\documentclass[11pt]{article}

\usepackage{theorem}

\theoremstyle{plain}

\usepackage{braket}
\usepackage{amssymb}
\usepackage{graphicx}
\usepackage[mathscr]{eucal}
\usepackage{titlesec}
\renewcommand{\thesubsection}{\thesection.\alph{subsection}}
\titleformat{\subsection}[runin]
{\normalfont\normalsize\bfseries}{\thesubsection}{0.5em}{}
\titlespacing*{\subsection} {0pt}{0.75ex plus 1ex minus .2ex}{0.5em}

{\theorembodyfont{\upshape} }
\newcommand{\boma}[1]{{\mbox{\boldmath $#1$} }}

\begin{document}
\newcommand{\vertiii}[1]{{\left\vert\kern-0.25ex\left\vert\kern-0.25ex\left\vert #1
    \right\vert\kern-0.25ex\right\vert\kern-0.25ex\right\vert}}
\newcommand{\binom}[2]{\left( \barray{c} #1 \\ #2 \farray \right)}
\newcommand{\uper}[1]{\stackrel{\barray{c} {~} \\ \mbox{\footnotesize{#1}}\farray}{\longrightarrow} }
\newcommand{\nop}[1]{ \|#1\|_{\piu} }
\newcommand{\no}[1]{ \|#1\| }
\newcommand{\nom}[1]{ \|#1\|_{\meno} }
\newcommand{\UD}[1]{e^{#1 \Delta}}
\newcommand{\bb}[1]{\mathbb{{#1}}}
\newcommand{\HO}[1]{\bb{H}^{{#1}}}
\newcommand{\Hz}[1]{\bb{H}^{{#1}}_{\zz}}
\newcommand{\Hs}[1]{\bb{H}^{{#1}}_{\ss}}
\newcommand{\Hg}[1]{\bb{H}^{{#1}}_{\gam}}
\newcommand{\HM}[1]{\bb{H}^{{#1}}_{\so}}
\newcommand{\vers}[1]{\widehat{#1}}
\def\sing{\mbox{\footnotesize{sing}}}
\def\x{\mathbf{x}}
\def\DD{\mathcal{D}}
\def\UU{\mathcal{U}}
\def\MM{\mathcal{M}}
\def\L{\mathscr{L}}
\def\T{\mathscr{T}}
\def\ellp{\lambda}
\def\tp{\tau}
\def\R{\mathscr{R}}
\def\Q{\mathscr{Q}}
\def\f{\mathscr{F}}
\def\Y{\mathscr{Y}}
\def\K{\mathscr{K}}
\def\V{\mathscr{V}}
\def\P{\mathscr{P}}
\def\J{\mathscr{J}}
\def\Ee{\mathcal{E}}
\def\Mis{\mbox{Meas}}
\def\Hs{W}
\def\Dom{\mbox{Dom}}
\def\bu{$\bullet$~}
\def\Hilb{\mathfrak{H}}
\def\Di{\mathfrak{D}}
\def\Ei{\mathfrak{E}}
\def\II{\mathscr{I}}
\def\KK{\mathscr{K}}
\def\Sii{\Sigma_{*}}
\def\Si{\Sigma}
\def\ti{s}
\def\lan{\lambda}
\def\C{C_0}
\def\tempo{\mathfrak{t}}
\def\sec{\mbox{sec}}
\def\m{\mbox{m}}
\def\Kg{\mbox{Kg}}
\def\Kilom{\mbox{Km}}
\def\Fi{\phi}
\def\Fip{\Fi'\,}
\def\rp{{r'}\,}
\def\hp{{h'}\,}
\def\rpp{{r''}\,}
\def\hpp{{h''}\,}
\def\te{\theta}
\def\ff{\varphi}
\def\xt{\dot{x}}
\def\xtt{\ddot{x}}
\def\yt{\dot{y}}
\def\ytt{\ddot{y}}
\def\qt{\dot{q}}
\def\ft{\dot{\phi}}
\def\ftt{\ddot{\phi}}
\def\at{\dot{a}}
\def\att{\ddot{a}}
\def\bt{\dot{b}}
\def\btt{\ddot{b}}
\def\barc{\bar{c}}
\def\einsu{\mathfrak{E}_{1}}
\def\einsd{\mathfrak{E}_{2}}
\def\conm{\mathfrak{G}_{(m)}}
\def\einsf{\mathfrak{G}_{\phi}}
\def\MK{\MM^d_K}
\def\tc{\tau}
\def\rom{\rho_m}
\def\Tm{{T^{(m)}}}
\def\Tf{{T^{(\phi)}}}
\def\gm{\eta}
\def\Sz{S_z}
\def\Bi{\mathscr{B}}
\def\xu{x^1}
\def\xd{x^2}
\def\xtp{{x^3}'}
\def\ki{k_i}
\def\ku{k_1}
\def\kd{k_2}
\def\kt{k_3}
\def\hu{h_1}
\def\hd{h_2}
\def\htt{h_3}
\def\ha{\widehat{a}}
\def\ak{\ha_k}
\def\ah{\ha_h}
\def\had{\ha^{\dagger}}
\def\akd{\had_k}
\def\ahd{\had_h}
\def\effe{F}
\def\Fk{\effe_k}
\def\Fh{\effe_h}
\def\Fkc{\overline{\Fk}}
\def\Fhc{\overline{\Fh}}
\def\fk{f_k}
\def\fh{f_h}
\def\fkc{\overline{\fk}}
\def\fhc{\overline{\fh}}
\def\bx{{\bf x}}
\def\Nab{\square}
\def\s{u}
\def\Fis{\Fi^{\s}}
\def\Tis{\Ti^{\s}}
\def\Aa{\widehat{A}}
\def\Bb{\widehat{B}}
\def\tu{\xi}
\def\ep{\varepsilon}
\def\iep{(-1,+\infty)}
\def\Do{\mathscr{E}}
\def\CA{\mathcal{C}}
\def\CB{\mathcal{D}}
\def\ca{c}
\def\cb{d}
\def\op{\,\mbox{\scriptsize{or}}\,}
\def\er{\epsilon}
\def\erd{\er_0}
\def\vk{\vers{k}}
\def\vh{\vers{h}}
\def\Qn{\mathfrak{K}_n}
\def\Ed{\hat{E}}
\def\um{u_{-}}
\def\up{u_{+}}
\def\el{t}
\def\em{z}
\def\uu{\lambda}
\def\dK{\delta {\mathscr K}}
\def\dG{\delta {\mathscr G}}
\def\DK{\Delta {\mathscr K}}
\def\DG{\Delta {\mathscr G}}
\def\Km{{\mathscr K}}
\def\Ll{\mathscr{L}}
\def\Hh{\mathscr{H}}
\def\Mm{{\mathscr M}}
\def\Nn{{\mathscr N}}
\def\Rr{{\mathscr R}}
\def\Gg{{\mathscr G}}
\def\Zz{Z}
\def\Ss{{\mathscr S}}
\def\Fe{{\mathscr F}}
\def\Ww{{\mathscr W}}
\def\we{\wedge}
\def\We{\bigwedge}
\def\dbar{\hat{d}}
\def\Cc{\mathscr{C}}
\def\SZ{\mathcal{S}}
\def\TZ{\mathfrak{S}}
\def\CQ{S}
\def\GQ{T}
\def\Ac{\overline{A}}
\def\Bc{\overline{B}}
\def\Xd{\Zd_{0 k} \cap \Bd_{0 k}(\ro)}
\def\Yd{\Zd_{0 k} \setminus \Bd_{0 k}(\ro) }
\def\comple{\scriptscriptstyle{\complessi}}
\def\nume{0.407}
\def\numerob{0.00724}
\def\deln{7/10}
\def\delnn{\dd{7 \over 10}}
\def\e{c}
\def\p{p}
\def\z{z}
\def\symd{{\mathfrak S}_d}
\def\Del{\delta}
\def\mmu{\hat{\mu}}
\def\rot{\mbox{rot}\,}
\def\curl{\mbox{curl}\,}
\def\XS{\boma{x}}
\def\TS{\boma{t}}
\def\Lam{\boma{\eta}}
\def\DS{\boma{\rho}}
\def\KS{\boma{k}}
\def\LS{\boma{\lambda}}
\def\PR{\boma{p}}
\def\VS{\boma{v}}
\def\ski{\! \! \! \! \! \! \! \! \! \! \! \! \! \!}
\def\h{L}
\def\EM{M}
\def\EMP{M'}
\def\E{E}
\def\FFf{\mathscr{F}}
\def\A{F}
\def\Xim{\Xi_{\meno}}
\def\Ximn{\Xi_{n-1}}
\def\om{\omega}
\def\Om{\Omega}
\def\Oma{\Om_a}
\def\Omp{\Om_{\infty}}
\def\Sim{\Sigm}
\def\Sip{\Delta \Sigm}
\def\Sigm{{\mathscr{S}}}
\def\Ki{{\mathscr{K}}}
\def\Hi{{\mathscr{H}}}
\def\zz{{\scriptscriptstyle{0}}}
\def\ss{{\scriptscriptstyle{\Sigma}}}
\def\gam{{\scriptscriptstyle{\Gamma}}}
\def\so{\ss \zz}
\def\Dz{\bb{\DD}'_{\zz}}
\def\Ds{\bb{\DD}'_{\ss}}
\def\Dsz{\bb{\DD}'_{\so}}
\def\Dg{\bb{\DD}'_{\gam}}
\def\Ls{\bb{L}^2_{\ss}}
\def\Lg{\bb{L}^2_{\gam}}
\def\bF{{\bb{V}}}
\def\Fz{\bF_{\zz}}
\def\Fs{\bF_\ss}
\def\Fg{\bF_\gam}
\def\Pre{P}
\def\UUU{{\mathcal U}}
\def\fiapp{\phi}
\def\PU{P1}
\def\PD{P2}
\def\PT{P3}
\def\PQ{P4}
\def\PC{P5}
\def\PS{P6}
\def\X{Q2}
\def\Xp{Q3}
\def\Vi{V}
\def\bVi{\bb{V}}
\def\Ks{\bb{\K}_\ss}
\def\Kz{\bb{\K}_0}
\def\KM{\bb{\K}_{\, \so}}
\def\HGG{\bb{H}^\G}
\def\HG{\bb{H}^\G_{\so}}
\def\EG{{\mathfrak{P}}^{\G}}
\def\G{G}
\def\de{\delta}
\def\esp{\sigma}
\def\dd{\displaystyle}
\def\LP{\mathfrak{L}}
\def\dive{\mbox{div}}
\def\la{\langle}
\def\ra{\rangle}
\def\um{u_{\meno}}
\def\uv{\mu_{\meno}}
\def\Fp{ {\textbf F_{\piu}} }
\def\Ff{ {\textbf F} }
\def\Fm{ {\textbf F_{\meno}} }
\def\piu{\scriptscriptstyle{+}}
\def\meno{\scriptscriptstyle{-}}
\def\omeno{\scriptscriptstyle{\ominus}}
\def\Tt{ {\mathscr T} }
\def\Xx{ {\textbf X} }
\def\Yy{ {\textbf Y} }
\def\VP{{\mbox{\tt VP}}}
\def\CP{{\mbox{\tt CP}}}
\def\cp{$\CP(f_0, t_0)\,$}
\def\cop{$\CP(f_0)\,$}
\def\copn{$\CP_n(f_0)\,$}
\def\vp{$\VP(f_0, t_0)\,$}
\def\vop{$\VP(f_0)\,$}
\def\vopn{$\VP_n(f_0)\,$}
\def\vopdue{$\VP_2(f_0)\,$}
\def\leqs{\leqslant}
\def\geqs{\geqslant}
\def\mat{{\frak g}}
\def\tG{t_{\scriptscriptstyle{G}}}
\def\tN{t_{\scriptscriptstyle{N}}}
\def\TK{t_{\scriptscriptstyle{K}}}
\def\CK{C_{\scriptscriptstyle{K}}}
\def\CN{C_{\scriptscriptstyle{N}}}
\def\CG{C_{\scriptscriptstyle{G}}}
\def\CCG{{\mathscr{C}}_{\scriptscriptstyle{G}}}
\def\tf{{\tt f}}
\def\ta{{\tt a}}
\def\tF{{\tt R}}
\def\TI{\tilde{I}}
\def\TJ{\tilde{J}}
\def\Lin{\mbox{Lin}}
\def\Hinfc{ H^{\infty}(\reali^d, \complessi) }
\def\Hnc{ H^{n}(\reali^d, \complessi) }
\def\Hmc{ H^{m}(\reali^d, \complessi) }
\def\Hac{ H^{a}(\reali^d, \complessi) }
\def\Dc{\DD(\reali^d, \complessi)}
\def\Dpc{\DD'(\reali^d, \complessi)}
\def\Sc{\SS(\reali^d, \complessi)}
\def\Spc{\SS'(\reali^d, \complessi)}
\def\Ldc{L^{2}(\reali^d, \complessi)}
\def\Lpc{L^{p}(\reali^d, \complessi)}
\def\Lqc{L^{q}(\reali^d, \complessi)}
\def\Lrc{L^{r}(\reali^d, \complessi)}
\def\Hinfr{ H^{\infty}(\reali^d, \reali) }
\def\Hnr{ H^{n}(\reali^d, \reali) }
\def\Hmr{ H^{m}(\reali^d, \reali) }
\def\Har{ H^{a}(\reali^d, \reali) }
\def\Dr{\DD(\reali^d, \reali)}
\def\Dpr{\DD'(\reali^d, \reali)}
\def\Sr{\SS(\reali^d, \reali)}
\def\Spr{\SS'(\reali^d, \reali)}
\def\Ldr{L^{2}(\reali^d, \reali)}
\def\Hinfk{ H^{\infty}(\reali^d, \KKK) }
\def\Hnk{ H^{n}(\reali^d, \KKK) }
\def\Hmk{ H^{m}(\reali^d, \KKK) }
\def\Hak{ H^{a}(\reali^d, \KKK) }
\def\Dk{\DD(\reali^d, \KKK)}
\def\Dpk{\DD'(\reali^d, \KKK)}
\def\Sk{\SS(\reali^d, \KKK)}
\def\Spk{\SS'(\reali^d, \KKK)}
\def\Ldk{L^{2}(\reali^d, \KKK)}
\def\Knb{K^{best}_n}
\def\sc{\cdot}
\def\k{\mbox{{\tt k}}}
\def\g{ {\textbf g} }
\def\QQQ{ {\textbf Q} }
\def\AAA{ {\textbf A} }
\def\gr{\mbox{gr}}
\def\sgr{\mbox{sgr}}
\def\loc{\mbox{loc}}
\def\PZ{{\Lambda}}
\def\PZAL{\mbox{P}^{0}_\alpha}
\def\epsilona{\epsilon^{\scriptscriptstyle{<}}}
\def\epsilonb{\epsilon^{\scriptscriptstyle{>}}}
\def\lgraffa{ \mbox{\Large $\{$ } \hskip -0.2cm}
\def\rgraffa{ \mbox{\Large $\}$ } }
\def\restriction{\upharpoonright}
\def\M{{\scriptscriptstyle{M}}}
\def\Fre{Fr\'echet~}
\def\ap{{\scriptscriptstyle{ap}}}
\def\fiap{\varphi_{\ap}}
\def\dfiap{{\dot \varphi}_{\ap}}
\def\DDD{ {\mathfrak D} }
\def\BBB{ {\textbf B} }
\def\EEE{ {\textbf E} }
\def\GGG{ {\textbf G} }
\def\TTT{ {\textbf T} }
\def\KKK{ {\textbf K} }
\def\HHH{ {\textbf K} }
\def\FFi{ {\bf \phi} }
\def\GGam{ {\bf \Gamma} }
\def\sc{ {\scriptstyle{\bullet} }}
\def\a{a}
\def\c{\kappa}
\def\parn{\par\noindent}
\def\teta{M}
\def\ro{\rho}
\def\al{\alpha}
\def\alc{\overline{\al}}
\def\dal{\mathfrak{a}}
\def\si{\sigma}
\def\be{\beta}
\def\dbe{\mathfrak{b}}
\def\bec{\overline{\be}}
\def\dbec{\overline{\dbe}}
\def\ga{\gamma}
\def\tet{\vartheta}
\def\teta{\theta}
\def\ch{\chi}
\def\et{\eta}
\def\complessi{\mathbb{C}}
\def\len{{\bf L}}
\def\reali{\mathbb{R}}
\def\Kap{\mathbb{K}}
\def\interi{{\bf Z}}
\def\Z{{\bf Z}}
\def\naturali{{\bf N}}
\def\To{ {\bf T} }
\def\Td{ {\To}^d }
\def\Tt{ {\To}^3 }
\def\Bd{B^d}
\def\Zd{ \interi^d }
\def\Zt{ \interi^3 }
\def\Zet{{\mathscr{Z}}}
\def\Ze{\Zet^d}
\def\Sfe{ {\bf S} }
\def\Sd{\Sfe^{d-1}}
\def\St{\Sfe^{2}}
\def\es{s}
\def\FF{\mathcal F}
\def\FFu{ {\textbf F_{1}} }
\def\FFd{ {\textbf F_{2}} }
\def\GG{{\mathcal G} }
\def\EE{{\mathcal E}}
\def\PP{{\mathcal P}}
\def\PPP{{\mathscr P}}
\def\PN{{\mathcal P}}
\def\PPN{{\mathscr P}}
\def\QQ{{\mathcal Q}}
\def\Np{{\hat{N}}}
\def\Lp{{\hat{L}}}
\def\Jp{{\hat{J}}}
\def\Vp{{\hat{V}}}
\def\Ep{{\hat{E}}}
\def\Gp{{\hat{G}}}
\def\Kp{{\hat{K}}}
\def\Ip{{\hat{I}}}
\def\Tp{{\hat{T}}}
\def\Mp{{\hat{M}}}
\def\La{\Lambda}
\def\Ga{\Gamma}
\def\Upsi{\Upsilon}
\def\Gam{\Gamma}
\def\Gag{{\check{\Gamma}}}
\def\Lap{{\hat{\Lambda}}}
\def\Upsig{{\check{\Upsilon}}}
\def\j{j}
\def\jp{{\hat{j}}}
\def\BB{{\mathcal B}}
\def\LL{{\mathcal L}}
\def\SS{{\mathcal S}}
\def\Dd{{\mathcal D}}
\def\VV{{\mathcal V}}
\def\WW{{\mathcal W}}
\def\OO{{\mathcal O}}
\def\RR{{\mathcal R}}
\def\TT{{\mathcal T}}
\def\AA{{\mathfrak A}}
\def\CC{{\mathcal C}}
\def\JJ{{\mathcal J}}
\def\NN{{\mathcal N}}
\def\HH{{\mathfrak H}}
\def\Ha{\hat{\HH}}
\def\XX{{\mathcal X}}
\def\XXX{{\mathscr X}}
\def\YY{{\mathcal Y}}
\def\ZZ{{\mathcal Z}}
\def\cir{{\scriptscriptstyle \circ}}
\def\circa{\thickapprox}
\def\vain{\rightarrow}
\def\parn{\par \noindent}
\def\salto{\vskip 0.2truecm \noindent}
\def\spazio{\vskip 0.5truecm \noindent}
\def\vs1{\vskip 1cm \noindent}
\def\fine{\hfill $\square$ \vskip 0.2cm \noindent}
\def\ffine{\hfill $\lozenge$ \vskip 0.2cm \noindent}
\newcommand{\rref}[1]{(\ref{#1})}
\def\beq{\begin{equation}}
\def\feq{\end{equation}}
\def\beqs{\begin{equation*}}
\def\feqs{\end{equation*}}
\def\beqq{\begin{eqnarray}}
\def\feqq{\end{eqnarray}}
\def\barray{\begin{array}}
\def\farray{\end{array}}

\makeatletter \@addtoreset{equation}{section}
\renewcommand{\theequation}{\thesection.\arabic{equation}}
\makeatother
\begin{titlepage}
{~} \vspace{0cm}
\begin{center}
{\huge On the linear instability of the Ellis-Bronnikov-Morris-Thorne wormhole}
\end{center}
\vspace{0.5truecm}
\begin{center}
{\large
Francesco Cremona$\,{}^a$, Francesca Pirotta$\,{}^b$, Livio Pizzocchero$\,{}^c$({\footnote{Corresponding author}})} \\
\vspace{0.5truecm}
${}^a$ Dipartimento di Matematica, Universit\`a di Milano\\
Via C. Saldini 50, I-20133 Milano, Italy \\
e--mail: francesco.cremona@unimi.it \\
\vspace{0.2truecm}
${}^b$ Dipartimento di Matematica, Universit\`a di Milano\\
Via C. Saldini 50, I-20133 Milano, Italy \\
e--mail: francesca.pirotta@yahoo.it \\
\vspace{0.2truecm}
${}^c$ Dipartimento di Matematica, Universit\`a di Milano\\
Via C. Saldini 50, I-20133 Milano, Italy\\
and Istituto Nazionale di Fisica Nucleare, Sezione di Milano, Italy \\
e--mail: livio.pizzocchero@unimi.it
\end{center}
\begin{abstract}
We consider the wormhole of Ellis, Bronnikov, Morris and Thorne (EBMT),
arising from Einstein's equations in presence of a
phantom scalar field. In this paper we propose a simplified
derivation of the linear instability of this system,
making comparisons with
previous works on this subject (and generalizations)
by Gonz\'{a}lez, Guzm\'{a}n, Sarbach,
Bronnikov, Fabris and Zhidenko.

\end{abstract}
\vspace{0.2cm} \noindent
\hfill \parn
\par \vspace{0.3truecm} \noindent \textbf{AMS subject classifications:} 83C15, 83C20, 83C25\,.
\par \vspace{0.3truecm} \noindent \textbf{PACS}: 04.20.Jb, 04.25.Nx\,.
\end{titlepage}
{~}
\vskip 0cm \noindent
\section{Introduction}
\label{intro}
Throughout this paper, indicating with $c,\hbar,G$
the speed of light, the reduced Planck constant and the
gravitational constant, we stipulate
\beq c = 1~, \qquad \hbar = 1~, \qquad \kappa := 8 \pi G~. \feq
We are interested in a well known wormhole; this is described by the
static spacetime metric
\beq d s^2 = - d t^2 + d \ell^2 + (a^2 + \ell^2) d\Om^2
\quad (-\infty < t, \ell <+ \infty)~ \label{thorne} \feq
where $d \Om^2 = d \te^2 + \sin^2 \te d \ff^2$ is the line
element of the unit spherical surface $S^2$ and $a$ is a positive
constant, with the dimension of a length.
For $\ell \vain \pm \infty$, $d s^2$ approaches the flat Minkowski metric $- d t^2 + d
\ell^2 + \ell^2 d\Om^2$. The region with $\ell \simeq 0$
represents the wormhole throat, of size $a$; this connects the
regions $\ell \gg a$, $\ell \ll -a$, representing two
asymptotically flat universes.
The spacetime geometry
\rref{thorne} received special attention in the classical 1988
paper by Morris and Thorne \cite{Mor}, considered as the origin of
modern investigations on wormholes.
\parn
Indeed, the line element
\rref{thorne} had appeared in the literature before \cite{Mor} (a
fact on which Thorne apologized in \cite{Div}). This spacetime
geometry was considered in a 1973 paper by Ellis \cite{Ell}, with
the denomination of ``drainhole'' (and with a somehow different
motivation, namely, to model an elementary particle); here the
metric \rref{thorne} was derived solving Einstein's equations in
presence of a massless scalar field $\phi$ minimally coupled to
gravity, after changing artificially the sign of the action
functional for $\phi$. Again in \cite{Ell}, the scalar field was
found to depend on $\ell$ with the law
\beq \phi = \sqrt{{2 \over \kappa}} \arctan{\ell \over a}~.
\label{field} \feq
Almost simultaneously to Ellis, Bronnikov \cite{Bron73} proposed a family
of scalar field solutions of Einstein's equations containing, as a
special case, the solution \rref{thorne} \rref{field}
({\footnote{\label{nota} The family of Bronnikov solutions depends
        on a ``mass'' parameter, which is zero in the case \rref{thorne}
        \rref{field}; see the recent paper of Yazadjiev \cite{Yaz} for
        an important uniqueness result on this family,
        and for a representation (in Eqs. (15)(16) of the cited article)
        very close to our notations.
        The mass-dependent generalization of the solution \rref{thorne} \rref{field}
        is never considered in the present work.}}). The scalar fields
considered by Ellis and Bronnikov, with an anomalous sign in (the
kinetic part of) their action functional, have become popular with
the denomination of phantom fields; their stress-energy violates
the usual conditions of positivity of the energy, thus mimicking
at the classical level a well known feature of quantum fields in
their vacuum states \cite{Hawk} \cite{Vis}.
\parn
In the rest of this paper we refer to the names or initials of the
previously mentioned authors and use the expressions ``EBMT
wormhole'', ``EBMT solution'' to indicate the phantom field
solution \rref{thorne} \rref{field} of Einstein's equations.
\parn
In this work we consider a perturbation of the solution
\rref{thorne} \rref{field} of the form \vfill \eject \noindent
{~}
\vskip -2cm \noindent
\par \noindent
\vbox{
    \beq d s^2 =  - d t^2 +
    \Bigg(1+ \ep \Q\Big({t \over a},{\ell \over a}\Big)\Bigg)^2 d \ell^2
    + \Bigg(\sqrt{a^2 + \ell^2}
    + \frac{\ep a^3}{a^2+ \ell^2}\R\Big({t \over a},{\ell \over a}\Big)\Bigg)^2 d\Om^2, \label{intro1} \feq
    \beq \phi = \sqrt{2 \over \kappa} \Bigg(\arctan{\ell \over a}+ \ep
    \Phi\Big({t \over a},{\ell \over a}\Big)\Bigg)
    \qquad (-\infty < t,\ell < + \infty)\, \label{intro2} \feq
}
where $\ep$ is a small real
parameter and $\Q,\R,\Phi$ are functions of the variables $s :=
t/a$, $x := \ell/a$, to be determined; Einstein's equations are
expanded to the first order in $\ep$, giving rise to a system of
linear equations for $\Q,\R,\Phi$.
\parn
Our handling of this
linear system produces in a simple way the general solution. As a
matter of fact, $\Q$ and $\Phi$ are represented explicitly as
functions of $\R$ (and of the initial data), and a ``master
equation'' is derived for $\R$; this has the form
\beq
\big(\partial_{s s} -\partial_{x x} + \V(x)\big) \R(s,x)= \J_0(x) + s \J_1(x)~, \qquad
\V(x) := - {3 \over (1 + x^2)^2}~, \label{msr}
\feq
where $\J_0, \J_1$ are source terms depending on the initial data
for the system. Since the operator $-\partial_{x x} + \V(x)$ has a
negative eigenvalue, Eq. \rref{msr} has solutions diverging
exponentially for large times; this suffices to infer the linear
instability of the EBMT solution.
\parn
Admittedly, the linear
instability of the EBMT system and of more general wormholes
supported by scalar fields has been stated previously in the literature, on the
grounds of suitably derived master equations for some
recombination of the perturbation components; therefore, it is
necessary to compare the present work with the previous papers on
this subject.
\parn
This comparison is performed in the forthcoming
subsections \ref{comp8} and \ref{comp9}; subsection \ref{secrem}
contains some remarks, and subsection \ref{org} concludes the present
introduction describing the organization of our work.
\subsection{Comparison with \cite{Proc}.}
\label{comp8}
When the results of the present work were derived,
we were not aware of the proceeding article \cite{Proc} by
Gonz\'{a}lez, Guzm\'{a}n and Sarbach while we had knowledge of
subsequent papers by the same authors, discussed hereafter
\cite{Gon} \cite{Gon2}; we were kindly informed about \cite{Proc}
by Professors Gonz\'{a}lez, Guzm\'{a}n and Sarbach, when we mailed
to them the first arXiv version (May 2018) of the present work.
\parn
Paper \cite{Proc} projects an elegant setting for the linear analysis of
the EBMT perturbed system, focusing on invariance features under
spacetime coordinate changes (gauge transformations)
infinitesimally close to the identity. The conclusion of the cited
article is that a suitable recombination $\chi$ of the
perturbation components fulfills
(in the notations of the present work)
$\big(\partial_{s s} -\partial_{x x} + \V(x)\big) \chi(s,x)=0$,
with $\V$ as in Eq. \rref{msr}; the same paper
proves that $-\partial_{x x} + \V(x)$ has a negative eigenvalue, a
fact yielding a virdict of linear instability. Unfortunately, the
discussion  of gauge transformations proposed in \cite{Proc}
contains some imprecision, which propagates to the formulation of
the linearized Einstein equations.
({\footnote{We acknowledge the
        authors of \cite{Proc} for an open and kind discussion on this
        subject. The analysis of infinitesimal gauge transformations in
        the cited paper fixes the attention on the radial coordinate
        ($\ell$ in our notations), and does not consider changes of the
        time coordinate $t$. The linearized Einstein equations of
        \cite{Proc} and the subsequent stability analysis are correct
        under the condition (not stated explicitly) that the field
        perturbation is zero. It is easy to check that the field
        perturbation always vanishes in a suitable coordinate system; of
        course, the choice of these distinguished coordinates breaks the
        desired gauge invariance of the overall setting.}})
\parn
In view of this, we think that a reconsideration of the perturbed
EBMT system in the linear approximation is not useless, even in
the simple approach proposed in the present work. Our analysis is
developed in a fixed gauge, defined requiring that the coefficient
of $-d t^2$ in the spacetime line element  be $1$ (on this, see
the comments accompanying our subsequent Eq. \rref{wormnonstat}).
As already mentioned, our master equation is written directly for
one component of the perturbation (the function $\R$ in
\rref{intro1}), with no need to form combinations with the other
components. The substantial nature of the large time divergences
arising from our computations is proved \textsl{a posteriori},
showing the impossibility to eliminate them via coordinate changes
(see the discussion in the last lines of subsection
\ref{ourinst}).
\subsection{Comparison with \cite{Gon} \cite{Gon2} \cite{Bron2011} \cite{Bron2018}.}
\label{comp9}
Paper \cite{Gon} by Gonz\'{a}lez, Guzm\'{a}n and
Sarbach considers Bronnikov's wormhole solution \cite{Bron73} of
the Einstein-scalar equations; as already indicated, the EBMT
system \rref{thorne} \rref{field} is a special case of this
solution . In \cite{Gon} the linear instability of the general
Bronnikov solution is derived via a two-steps construction, that
we now describe briefly. The first step is the reduction of the
linearized Einstein equations to a scalar master equation where
the unknown is a suitable recombination of the perturbation
components, here indicated with $\chi_{\sing}$. The potential
$\V_{\sing}$ in this master equation is singular at the wormhole
throat; other singularities, again located at the throat, affect a
source term appearing in the same equation and the very definition
of the recombination $\chi_{\sing}$. The second step in the
construction of \cite{Gon} removes the singularities by a clever
strategy: the idea is to apply to $\chi_{\sing}$ a suitable first
order differential operator, so as to obtain a function $\chi$
fulfilling a regular master equation. This is in fact possible if
one knows a static solution of the singular master equation; the
static solution determines the transformation relating
$\chi_{\sing}$ and $\chi$.
In our notations the final, regular
master equation reads $(\partial_{s s} - \partial_{x x} + \V(x))
\chi(s,x)=0$, where $\V$ is a nowhere singular potential; the
authors of \cite{Gon} show that the operator $-\partial_{x x} +
\V(x)$ has a negative eigenvalue, a fact implying the linear
instability of the Bronnikov solution.
\parn
In the special EBMT case, the regular potential $\V$
coincides with the function in
Eq. \rref{msr}; however, as indicated before, in the EBMT case a
regular master equation can be derived in a direct way with no
need to use the previous two-steps construction.
\parn
Paper \cite{Gon} has a companion work by the same authors \cite{Gon2}
where the exact, nonlinear Einstein equations for the perturbed
Bronnikov solution are treated numerically, providing evidence
that the initial perturbation produces a rapid growth of
the wormhole's throat or a collapse to a black hole.
(A numerical analysis of the exact, perturbed Einstein equations
is also given in the second
half of \cite{Proc} for the special EBMT case).
Admittedly, this issue is beyond the aims of the present
work.
\parn
Returning to the linear stability analysis, let us
point out that the two-steps approach (a singular master equation,
a subsequent regularization) has been extended by Bronnikov,
Fabris and Zhidenko \cite{Bron2011} to the whole class of static,
radially symmetric scalar field solutions of Einstein's equations
with throats (including cases with an external potential for the
scalar field). Let us also mention a very recent paper of
Bronnikov \cite{Bron2018}, an excellent review about wormholes and
black holes supported by scalar fields that considers, amongst
else, the two-steps approach to linear stability problems.
\subsection{{Some remarks.}}
\label{secrem}
For completeness, let us add some comments
on two issues which have partial relations
with the present work, but fall outside its scope.
\par \noindent
(i) A phantom scalar field is not the unique
source producing the metric \rref{thorne} via
Einstein's equations. Another source
has been considered by
Shatskii, Novikov and Kardashev
\cite{Nov}: this consists of a ``phantom'' fluid
(with negative mass-energy density) and of an electromagnetic
field. Of course, this alternative source
requires a separate analysis for the stability
problem. Bronnikov,
Lipatova, Novikov and Shatskiy
\cite{Bronfl} have shown that,
assuming a non conventional equation
of the state for the fluid, the system is linearly stable
under radially symmetric and axial perturbations;
the same authors have conjectured the linear
stability under arbitrary perturbations.
\par \noindent
(ii) The stability analysis for a wormhole
supported by some kind of field is a subject
that differs from the study of the wave
equation for a test scalar or electromagnetic field in the
background of a given static wormhole (i.e., with a fixed
spacetime metric). Investigations along this second line are
currently very active and produced a lot of interesting results,
ranging from wave scattering theory in the wormhole background to
the reconstruction of the wormhole shape by inverse scattering
techniques: see, in particular, the recent work by Konoplya
\cite{Kon} and its bibliography.
Due to some similarities
between the terminologies employed in the two areas
of linear stability and test field analysis, it is not
useless to mention the existence of conceptual differences.
\subsection{Organization of the paper.} \label{org}
Making reference to subsection \ref{comp8} for the motivations of
the present work, let us briefly outline its organization. Section
\ref{basic} reviews some basic facts on (ordinary and) phantom
scalar fields minimally coupled to gravity, and on Einstein's
equations for such systems with the assumption of radial symmetry;
the EBMT solution \rref{thorne} \rref{field} is presented as a
static solution of these equations.
\parn
Section \ref{linpert} is
the core of the paper. In subsections \ref{ourpert}-\ref{uourlin}
we perturb the EBMT solution as in Eqs. \rref{intro1}
\rref{intro2}, and linearize the corresponding Einstein equations.
In subsections \ref{ourfi}-\ref{ourmaster} we express all
perturbation components in terms of $\R$, and derive a master
equation for this component. In the final subsections
\ref{ourspec}-\ref{ourinst} we write down the general solution of
our master equation (hence, of the linearized Einstein equations);
we show that there are solutions diverging exponentially for large
times, and that such divergences cannot be eliminated by
coordinate changes. In \ref{ourspec}-\ref{ourinst} we also take
the occasion to set up a rigorous functional-analytic framework
for the master equation, based on the language of Sobolev spaces.
Concerning functional aspects, let us point out that the term
``smooth'' often used in the sequel always means $C^\infty$.
\newpage
\section{Some basic facts}
\label{basic}
\subsection{Gravitation and scalar fields.}
\label{2a}
In a four-dimensional spacetime, we consider a
gravitational field minimally coupled to a real scalar field
$\phi$ with a vanishing field self-potential, i.e., a real scalar
field with zero mass and no self-interaction. This system is
described by the action functional
\beq S[g_{\mu \nu}, \phi] :=
\int \left(\,{R \over 2 \kappa}
- {\si \over 2} \partial^{\mu} \phi \, \partial_{\mu} \phi\, \right) d v
\feq
where: $g_{\mu \nu}$ is the spacetime metric (of course used
to raise and lower indices); $R$ and $d v$ are the scalar curvature and the volume
element corresponding to this metric
($d v = \sqrt{|\det(g_{\mu \nu})|} \prod^{\,}_{\lan} d x^\lan$
in any spacetime coordinate system $(x^\lan)$);
$\si := 1$ for an ordinary field, $\si := -1$
for a phantom field. Both the metric and the scalar field are
always assumed to be smooth.
\parn
The stationarity condition $\delta S/\delta g_{\mu \nu}=0$ gives
Einstein's equations
\beq R_{\mu \nu} - {1 \over 2} g_{\mu \nu} R
= \kappa T_{\mu \nu} \label{einseq} \feq
where the right side contains the field
stress-energy tensor ({\footnote{As well known, Einstein's equations have the
        equivalent form $R_{\mu \nu} =\kappa \big(T_{\mu \nu}-{1 \over
            2}g_{\mu \nu} T\big)= \sigma
        \kappa\partial_\mu\phi\partial_\nu\phi $, used in many of the
        previously cited works. For our manipulations on the linearized
        equations, the form \rref{einseq} is more convenient. }})
\beq T_{\mu \nu} :=
\si \left(\, \partial_{\mu} \phi \, \partial_{\nu} \phi
- {1 \over 2} g_{\mu \nu}
\partial^{\lan} \phi \, \partial_{\lan} \phi  \right) \label{tens} ~.\feq
The stationarity condition $\delta S/\delta \phi=0$ gives the field
equation
\beq \square \phi =0 ~\label{fieldeq} \feq
where $\square := \nabla_{\mu} \nabla^\mu$ and
$\nabla_{\mu}$ is the covariant
derivative induced by the metric $g_{\mu \nu}$
($\nabla_\mu = \partial_\mu$ on scalar functions, like $\phi$).
\parn
Indeed, Einstein's equations \rref{einseq} imply the field
equation \rref{fieldeq}. In fact, Einstein's equations and the
contracted Bianchi identity give $\nabla_{\mu} T^{\mu}_{~\nu} =0$
and, on the other hand, the definition \rref{tens} implies
$\nabla_{\mu} T^{\mu}_{~\nu} = \si \,(\square \phi)\,
\partial_{\nu} \phi$; thus
(\ref{einseq}) $\Rightarrow$ $(\square \phi)\, \partial_{\nu} \phi=0$.
Refining these considerations, one
obtains that
({\footnote{Here is a derivation of \rref{thus}. Let
        us assume Einstein's equations \rref{einseq}; then
        $(\square \phi)\, \partial_{\nu} \phi=0$ or, in index-free notation,
        $(\square \phi)\, d \phi=0$ where $d$ is the usual differential.
        Denoting with $\MM$ the spacetime, let us introduce the open set
        $\DD := \{ \x \in \MM~|~(d \phi)(\x) \neq 0 \}$. Of course
        $\square \phi=0$ on $\DD$; hereafter we show that $\square \phi=0$
        even on the complementary set $\MM \setminus \DD$. In fact, let
        $\x \in \MM \setminus \DD$; then $\x$ belongs to the frontier
        $\partial \DD$, or $\x$ is an inner point of $\MM \setminus \DD$.
        If $\x \in \partial \DD$ each neighborhood of $\x$ contains a
        point $\x' \in \DD$, for which $(\square \phi)(\x')=0$; so, by
        continuity, $(\square \phi)(\x)=0$. If $\x$ is an inner point of
        $\MM \setminus \DD$, let us choose an open connected neighborhood
        $\UU$ of $\x$ such that $\UU \subset \MM \setminus \DD$; then $d
        \phi=0$ on $\UU$, whence $\phi=$ constant on $\UU$ and,
        consequently, $\square \phi=0$ on $\UU$.}})
\beq
\mbox{(\ref{einseq})}~ \Longrightarrow~ \square \phi=0~.
\label{thus} \feq
\\
\subsection{The radially symmetric case.}
Now, let us consider a spacetime with line element $d
s^2$ and a scalar field $\phi$, where
\beq
d s^2 =  - d t^2 + q^2(t,\ell) d \ell^2 + r^2(t,\ell) d\Om^2,~~
\phi = \Fi(t,\ell) \quad (-\infty < t, \ell < + \infty).
\label{wormnonstat}
\feq
In the above $d \Om^2 = d \te^2 + \sin^2 \te d \ff^2$ denotes
(again) the line element of the unit spherical surface $S^2$ ($0 <
\te < \pi$, $0 < \ff < 2 \pi$) and $q(\,,\,)>0$, $r(\,,\,)>0$,
$\phi(\,,\,)$ are
smooth functions.
\parn
Let us mention that a line element of the seemingly more general
form $d s^2 = - h(t,\ell)^2 d t^2 + q^2(t,\ell) d \ell^2 +
r^2(t,\ell) d\Om^2$ can be reduced (at least locally) to the form
in \rref{wormnonstat}, with $h=1$, performing a suitable
coordinate change $(t,\ell) \to (t', \ell')$
({\footnote{This follows, e.g., from the general
        discussion of \cite{Lan}, $\S 97$
        on synchronous coordinate systems on arbitrary spacetimes.}}).
\parn
From here to the end of the paper, we make systematic reference to
Eq.\,\rref{wormnonstat} and to the coordinate system
\beq (t,\ell,\te,\ff) \equiv (x^\mu)_{\mu=t,\ell,\te,\ff}~. \feq
The
configuration that we are considering is radially symmetric. For the metric $g_{\mu \nu}$ and the field $\phi$
described by \rref{wormnonstat}, the only independent Einstein
equations are those corresponding to the choices $(\mu,\nu) =
(t,t),(t,\ell),(\ell,\ell),(\te,\te)$ that read, respectively
({\footnote{$R_{\mu \nu} - {1 \over 2} g_{\mu \nu} R$ is as
        follows: it equals the left hand side of Eq.
        \rref{tt}, \rref{ellt}, \rref{ellell}, \rref{tete}, respectively,
        for $(\mu,\nu) = (t,t),
        (t,\ell)\,\mbox{or}\,(\ell,t),(\ell,\ell),(\te,\te)$; it equals
        $\sin^2\te$ $\times$ the left hand side of Eq. \rref{tete}, for
        $(\mu,\nu) =(\ff,\ff)$; it vanishes for all the other choices of
        $(\mu,\nu)$. One can make similar statements for $\kappa
        T_{\mu\nu}$, using the right hand sides of Eqs.
        (\ref{tt}-\ref{tete}). }}):
\par \noindent
\vbox{
    \beq
    \frac{1}{r^2 } +\frac{2q_{t}r_{t}}{q r} + \frac{2q_{\ell}\,r_{\ell }}{q^3 r}
    + \frac{r_{t}^2}{r^2}   - \frac{r_{\ell}^2}{q^2 r^2}
    - \frac{2 r_{\ell \ell}}{q^2 r}
    =\frac{\sigma \kappa}{2}\Bigg(\phi_t^2+\frac{\phi_\ell^2}{q^2}\Bigg), \label{tt} \feq
    \beq
    \frac{2 q_{t}r_{\ell }}{q r}- \frac{2 r_{t \ell}}{r} =\sigma
    \kappa\phi_t \phi_\ell~, \label{ellt} \feq
    \beq - \frac{q^2}{r^2}
    - \frac{q^2 r_{t}^2}{r^2} + \frac{r^2_{\ell}}{r^2} - \frac{2 q^2 r_{tt}}{r}
    =\frac{\sigma \kappa}{2}\Bigg(q^2\phi_t^2+\phi_\ell^2\Bigg) ,\label{ellell} \feq
    \beq -\frac{q_{t} r r_{t}}{q} -\frac{q_{\ell } \, r r_{\ell }}{q^3} -
    \frac{q_{tt} r^2}{q} -r r_{tt}
    + \frac{r r_{\ell \ell }}{q^2}
    =\frac{\sigma \kappa r^2}{2}\Bigg({\phi_t^2} -\frac{\phi_\ell^2}{q^2}\Bigg) \label{tete}
    \feq
}
(here and in the sequel, subscripts like ${}_\ell$ or ${}_t$ are
used to indicate derivatives). The field equation $\square \phi=0$
will not even be written since, according to \rref{thus}, it is a
consequence of Eqs. (\ref{tt}-\ref{tete}). For future use we
record the explicit expression of the scalar curvature for the
metric \rref{wormnonstat}, which is as follows:
\beq
R = {2 \over r^2} + {4 q_t r_t\over q r} + {4
    q_{\ell}r_{\ell}\over q^3 r}
+ \frac{2 r_t^2}{r^2} - \frac{2 r_{\ell}^2}{q^2 r^2}
+ {2 q_{tt} \over q} + {4 r_{tt} \over r} -{4 r_{{\ell}{\ell}}
    \over q^2 r}~. \label{scal}
\feq
\newpage
\subsection{The EBMT wormhole \cite{Mor,Ell,Bron73}.}
\label{2c}
This corresponds to the following static solution of
the Einstein equations (\ref{tt}-\ref{tete}):
$$
q(\ell):=1,~~r(\ell) := \sqrt{a^2 + \ell^2}, $$
\beq \sigma := -1~\mbox{(phantom field)},
~~ \phi(\ell) := \sqrt{2 \over \kappa}\, \arctan {\ell \over a}
\quad
(-\infty < \ell < + \infty), \label{solellis}
\feq
where $a > 0$ is a parameter, with the dimension of a length. The
line element $d s^2$ corresponding to \rref{solellis} has the form
\rref{thorne}; it describes a traversable wormhole with a throat
of size $\inf_{\ell} r(\ell) = a$. In the present case, Eq.
\rref{scal} for the scalar curvature gives
\beq R = - {2 a^2 \over (a^2 + \ell^2)^2}~. \feq
\\
\section{Linear instability of the EBMT wormhole: a simplified derivation.}
\label{linpert}
From here to the end of the paper $\phi$ is a phantom scalar
field, i.e., \beq \si := -1~. \feq
\subsection{Radial perturbations of the EBMT solution.}
\label{ourpert}
We consider a line element $ds^2$ and a scalar
field $\phi$ as in \rref{wormnonstat}, with
$$ q(t,\ell):=1+ \ep \Q\Bigg({t \over a},{\ell \over a}\Bigg), \qquad
r(t,\ell):=\sqrt{a^2 + \ell^2}
+ \frac{\ep a^3}{a^2+ \ell^2}\R\Bigg({t \over a},{\ell \over a}\Bigg), $$
\beq \phi(t,\ell):=\sqrt{2 \over \kappa} \left(\arctan{\ell \over a}
+ \ep \Phi\Bigg({t \over a},{\ell \over a}\Bigg)\right)~;
\label{pert}
\feq
here $\ep \in \reali$ is a small dimensionless parameter (that we
ultimately send to zero) and $\Q,\R,\Phi: \reali\times\reali \vain
\reali$ are smooth dimensionless functions, to be determined;
these depend on the variables \beq \ti := t/a~, \qquad x := \ell/a
\feq which are dimensionless in our units with $c=1$. The factor
$a^3/(a^2 + \ell^2)$ multiplying $\R$ in Eq. \rref{pert} will
simplify our subsequent calculations. (Note the equivalence
between Eq. \rref{pert} and Eqs. \rref{intro1}\rref{intro2} of the
Introduction).
\subsection{Linearizing Einstein's equations
(and the scalar curvature).}
\label{uourlin}
Let us substitute the
expressions \rref{pert} into Einstein's equations
(\ref{tt}-\ref{tete}) and expand them up to the first order in
$\ep$. Of course, these equations are satisfied to the zeroth
order in $\ep$, corresponding to the EBMT solution; moreover, Eqs.
(\ref{tt}-\ref{tete}) hold to the first order in $\ep$ if and only
if we have, respectively:
\beq
\Q+ x \Q_x
+ \frac{2(1-2x ^2) \R}{(1+x ^2)^\frac{5}{2}}
+\frac{3x \R_x}{(1+x ^2)^\frac{3}{2}}
-\frac{\R_{x x}}{\sqrt{1+x ^2}}
+\Phi_x =0 \, , \label{lit} \feq
\beq
x \Q_\ti + \frac{2x \R_\ti}{(1+x^2)^\frac{3}{2}}
-\frac{\R_{\ti x}}{\sqrt{1+x^2}} +\Phi_\ti =0 \, , \label{liellet} \feq
\beq \Q -\frac{(1-2x^2) \R}{(1+x^2)^\frac{5}{2}}
-\frac{x \R_x}{(1+x^2)^\frac{3}{2}}
+\frac{\R_{\ti\ti}}{\sqrt{1+x^2}}-\Phi_x =0 \, ,
\label{lielle} \feq
\beq
x \Q_x + (1+x^2)\Q_{\ti\ti}
+ \frac{3(1-2x^2) \R}{(1+x^2)^\frac{5}{2}}
+ \frac{4x \R_{x}}{(1+x^2)^\frac{3}{2}}
+ \frac{\R_{\ti \ti}-\R_{x x}}{\sqrt{1+x^2}} + 2\Phi_x=0 \, .
\label{lite}
\feq
By obvious considerations based on \rref{thus}, the linearized
Einstein's equations (\ref{lit}-\ref{lite}) ensure the field
equation $\square \phi=0$ to hold as well up to the first order in
$\ep$. For future use we also write down the first order expansion
of the scalar curvature \rref{scal}, which is as follows:
\par \noindent
\vbox{
    $$
    R =  - {2 a^2 \over (a^2 + \ell^2)^2}
    + \frac{2\ep}{a^2} \K\left({t \over a},{\ell \over a}\right) + O(\ep^2)~, $$
    \beq \K :=  {2 (2 + x^2) \Q  \over (1 + x^2)^2} + {x \Q_x  \over 1 + x^2}  +  \Q_{\ti \ti}
    + {4 (1 - 2 x^2) \R  \over (1 + x^2)^{7/2}}
    + {6x \R_x  \over (1 + x^2)^{5/2}} + 2 {\R_{\ti \ti} - \R_{x x}
        \over (1 + x^2)^{3/2}}\, . \label{scalesp}
    \feq
}
The subsequent subsections contain our treatment of the linearized
equations (\ref{lit}-\ref{lite}); a distinctive feature of our
approach is that we first determine the field perturbation $\Phi$,
as described hereafter.
\subsection{Finding $\boma{\Phi}$.}
\label{ourfi}
Integrating with respect to $\ti$, we see that Eq.
\rref{liellet} holds if and only if
$\Phi(\ti,x)= -x \Q(\ti,x) - 2x \R(\ti,x)/(1+x ^2)^\frac{3}{2}
+\R_x(\ti,x)/\sqrt{1+x^2}+ \mathscr{C}(x)$
where $\mathscr{C}: \reali \vain \reali$ is a
smooth function. Inserting this espression for $\Phi$ into Eq.
\rref{lit}, we see that the latter holds if and only if
$\mathscr{C}$ is constant. Summing up: Eqs.
\rref{lit}\rref{liellet} hold if and only if
\beq \label{psia}
\Phi(\ti,x)=
- x \Q(\ti,x) -\frac{2 x \R(\ti,x)}{(1+x ^2)^\frac{3}{2}}
+\frac{\R_{x }(\ti,x)}{\sqrt{1+x ^2}} + \C \, , \feq
where $\C \in \reali$ is a constant. The value of $\C$ is
immaterial (note that $\Phi$ appears in the linearized equations
(\ref{lit}-\ref{lite}) only through its derivatives; the same can
be said for $\phi$ in the exact equations (\ref{tt}-\ref{tete})).
\subsection{$\boma{\Q}$, $\boma{\Phi}$ as functions of $\boma{\R}$ and of the initial data.}
\label{ourqr}
Now we are left with Eqs. \rref{lielle} \rref{lite};
evidently, this pair is equivalent to the pair formed by Eq
\rref{lielle} and by Eq. \rref{lielle} + Eq. \rref{lite} (side by
side sum). The combination \rref{lielle} + \rref{lite} is reduced,
after substituting the expression \rref{psia} for $\Phi$,
to the equation
\beq \label{A'} \Q_{\ti\ti}
+ \frac{2 \R_{\ti\ti}}{(1+x ^2)^\frac{3}{2}}  =0~; \feq
this holds if and only if
$\Q(\ti,x) = -2 \R(\ti,x)/(1+x ^2)^\frac{3}{2}$
$+$ a function with vanishing $\ti\ti$-derivative, i.e.,
\beq \label{q} \Q(\ti,x)=-\frac{2 \R(\ti,x)}{(1+x^2)^\frac{3}{2}}
+ \P_0(x) + \ti\, \P_1(x) \feq
where $\P_0,\P_1 : \reali \vain \reali$ are smooth functions;
these are closely related to the set of initial data
\beq
{~}
\hspace{-0.35cm}
\Q_0(x) := \Q(0,x), \R_0(x) := \R(0,x),
\Q_{1}(x) := \Q_\ti(0,x), \R_{1}(x) := \R_\ti(0,x), \label{qri} \feq
since \rref{q} implies
\beq \label{ui} \P_i(x) = \Q_i(x)
+ \frac{2 \R_i(x)}{(1+x^2)^\frac{3}{2}} \quad (i=0,1)~.
\feq
Returning to Eq. \rref{psia} for $\Phi$, and substituting
therein Eq. \rref{q} for $\Q$, we obtain for the
field perturbation the final expression
\beq
\Phi(\ti,x)=\frac{\R_{x}(\ti,x)}{\sqrt{1+x^2}}
- x \, \Bigg(\P_0(x) + \ti \, \P_1(x) \Bigg)+\C\, \qquad
(\P_i ~\mbox{as in \rref{ui}})~. \label{psi}
\feq
\subsection{A master equation for $\boma{\R}$.}
\label{ourmaster}
We finally substitute the expressions \rref{q}
\rref{psi} for $\Q,\Phi$ into \rref{lielle}; the equation obtained
in this way holds if and only if
\beq (\R_{\ti\ti} + \HH \,\R)(\ti,x) = \J_0(x) + \ti \, \J_1(x)~,
\label{eqr} \feq
where
\beq \HH := - {d \over d x^2} + \V \, , \quad
\V(x) := -\frac{3}{(1+x^2)^2} \quad (x \in \reali)\,, \label{hh} \feq
\beq
\label{ji}
\J_i(x) := - \sqrt{1+x^2} \Bigg(2 \,\P_i(x) + x \,\P_{i,x}(x)\Bigg)
\quad (i=0,1,~\P_i ~\mbox{as in \rref{ui}}).\feq
$\HH$ is, formally, a
Schr\"odinger type operator in space dimension $1$ with potential
$\V$; the functions $\J_i$ are fully determined by the functions
$\P_i$ or, due to \rref{ui}, by the initial data $\Q_i$, $\R_i$
($i=0,1$). Eq. \rref{eqr} is our master equation; it is a
wave-type equation for $\R$ with a source term $\J_0(x) + \ti \,\J_1(x)$.
\subsection{Spectral analysis tools to solve the master equation.}
\label{ourspec}
The solution of Eq. \rref{eqr} is reduced to the
spectral analysis of the operator $\HH$ defined by \rref{hh}, in a
convenient Hilbertian framework; in view of this, from now on the
derivative $d^2/ d x^2$ appearing therein will be intended in the
most general sense, i.e., in the sense of the Schwartz
distributions theory \cite{Rud}. From the general theory of
Schr\"odinger operators on the real line with smooth potentials
vanishing at infinity \cite{Ber}, one infers the following
statements (i-iii):
\begin{itemize}
    \item[(i)] Consider the Hilbert space
    $L^2(\reali,\complessi)$ of complex valued, square integrable
    functions on $\reali$; let $\Ha$ denote the restriction of $\HH$
    to the domain
    $\{ \f \in L^2(\reali,\complessi)~|~\HH \f \in L^2(\reali,\complessi) \}$;
    then, $\Ha$ is a selfadjoint operator in $L^2(\reali,\complessi)$.
    \item[(ii)] The discrete spectrum $\sigma_d(\Ha)$ consists of finitely
    many, negative eigenvalues; the continuous spectrum
    $\sigma_c(\Ha)$ coincides with $[0,+\infty)$. Any eigenvalue $-E
    \in \sigma_d(\Ha)$ has an associated space of (smooth) square
    integrable eigenfunctions, of dimension $1$. Every point $W \in
    (0,+\infty)$ has an associated, 2-dimensional space of
    ``generalized'' eigenfunctions: these are (smooth) functions $\Y$
    which fulfill $\HH \Y = W \Y$ but do not belong to
    $L^2(\reali,\complessi)$.
    \item[(iii)] Choosing appropriately a
    normalized eigenfunction $\Y_{-E}$ for each eigenvalue $-E \in
    \sigma_d(\Ha)$ and two generalized eigenfunctions $\Y^{j}_{W}$
    ($j=1,2$) for each $W \in (0,+\infty)$, one can build a
    ``generalized'' orthonormal basis for $L^2(\reali, \complessi)$.
    These choices can be made so that all the previous eigenfunctions
    are real valued.
\end{itemize}
We will now profit from the analysis
already performed in \cite{Proc} \cite{Gon} for the operator
$\Ha$, resting on specific features of its potential $\V$. In
\cite{Proc}, it is shown that $\Ha$ has at least one (necessarily
negative) eigenvalue; in \cite{Gon} it is proved that the discrete
spectrum of $\Ha$ consists of exactly one eigenvalue, and a
numerical estimate is given for it:
({\footnote{Paper \cite{Gon}
        does not report directly the value of $E$ but, rather, the
        dimensionless ``unstability time'' $T := 1/\sqrt{E}$. For this
        quantity it is stated that $T \simeq 0.846$ (see Table 1 of the
        cited work); this implies for $E$ the estimate in \rref{defe}.}})
\beq \sigma_d(\Ha) = \{ - E \}~,
\qquad E \simeq 1.40~. \label{defe} \feq
According to (iii), we have a generalized
orthonormal basis formed by a normalized, real valued
eigenfunction $\Y_{-E}$ and by a pair of generalized, non square
integrable real valued eigenfunctions $\Y^{j}_{W}$ ($j=1,2$) for
each $W \in (0,+\infty)$. It should be noted that $\Y_{-E}$ is an
even function: $\Y_{-E}(-x) = \Y_{-E}(x)$; this reflects a general
result on the eigenfunction for the minimum eigenvalue of a
Schr\"odinger operator $- d^2/ d x^2 + \V$ with an even potential
$\V$.
\parn
From now on $\Kap := \reali$ or $\complessi$; we consider the
space $L^2(\reali, \Kap)$ of square integrable functions from
$\reali$ to $\Kap$. For each $\f \in L^2(\reali, \Kap)$ we have
(intending suitably all the integrals that follow \cite{Ber})
\beq
\f(x) = \la \Y_{-E}| \f \ra \Y_{-E}(x) + \sum_{j=1,2}
\int_{0}^{+\infty} d W \la \Y^{j}_{W}| \f \ra
\Y^{j}_{W}(x)~, \label{espef}
\feq\\
where $\la \Y | \f \ra := \int_{\reali} d x \,\Y(x) \f(x) \in
\Kap$ for $\Y = \Y_{-E}, \Y^{j}_{W}$.
Moreover, let $\|~\|$ denote the norm of
$L^2(\reali,\Kap)$ defined by
$\| \f \|^2 = \int_{\reali} |\f(x)|^2 d x$;
then, we have the representation
$\| \f \|^2 = |\la \Y_{-E}| \f \ra|^2 + \sum_{j=1,2}
\int_{0}^{+\infty} d W |\la \Y^{j}_{W}| \f \ra|^2$.
If $\f$ and $\HH \f$ are both in $L^2(\reali, \Kap)$, one also
has $\la \Y_{-E} | \HH \f \ra = - E \la \Y_{-E} | \f \ra$ and
$\la \Y^{j}_{W}| \HH \f \ra = W \la \Y^{j}_{W}| \f \ra$ for $W > 0$,
$j=1,2$.
\parn
To go on, one can introduce the function space
\beq \Ee(\reali,\Kap) := \{ \f~|~\f, \HH \,\f, \HH^2 \f...
\in L^2(\reali,\Kap) \}~, \feq
which is a Fr\'echet space \cite{Rud}
with the countably many norms $\f \mapsto \| \f \|$, $\| \HH \f
\|$, $\| \HH^2 \f \|$,...~; note that $\Y_{-E} \in \Ee(\reali,\Kap)$.
By means of some Sobolev imbeddings (see again
\cite{Rud}, Theorem 7.25), one shows that
$\Ee(\reali,\Kap) = \{ \f \in C^\infty(\reali,\Kap)~|~
\f, \f_{x}, \f_{xx},... \in L^2(\reali,\Kap)\}$
and that the previous family of norms is
topologically equivalent to the family of (semi-)norms
$\f \mapsto \| \f \|, \| \f_{x} \|, \| \f_{x x} \|,...$\,.
\subsection{Solving the master equation for $\boma{\R}$; conclusions for the linearized
Einstein equations.}
\label{oursol}
Let us keep all notations of subsection \ref{ourspec};
in particular, $\Y_{- E}$ and $\Y_{W}^j$
are the real valued eigenfunctions in item (iii) therein.
\parn
Assume that $\Q,\R,\Phi \in C^\infty(\reali^2,\reali)$ are
solutions of the linearized Einstein equations
(\ref{lit}-\ref{lite}); then, we have Eqs. \rref{q}\rref{psi} for
$\Q, \Phi$ and the master equation \rref{eqr} for $\R$ (all of
them involving the initial data $\Q_i, \R_i$ of Eq. \rref{qri}
through Eqs. \rref{ui}\rref{ji}). In addition, assume that:
\begin{itemize}
    \item[($\alpha$)] $\J_i \in \Ee(\reali,\reali)$ for $i=1,2$ (this is, in
    fact, a condition about the data $\Q_i$ and $\R_i$ defining $\J_i$
    via Eqs. \rref{ui} \rref{ji}).
    \parn
    \item[($\beta$)] For each $\ti \in \reali$, the function
    $\R(\ti,\cdot) : x \mapsto \R(\ti,x)$ is in $\Ee(\reali,\reali)$
    and the mapping $\ti \mapsto \R(\ti,\cdot)$ is $C^\infty$
    from $\reali \ni \ti$ to the space $\Ee(\reali,\reali)$.
\end{itemize}
Then, at each ``time'' $\ti$, we have an expansion of the form
\rref{espef} for $\f := \R(\ti,\cdot)$. It is inferred from
\rref{eqr} that
\hbox{$({d^2/d \ti^2} - E) \la \Y_{-E}| \R(\ti,\cdot) \ra$}
\hbox{$= \la \Y_{-E}| \J_0 \ra$} \hbox{$+ \ti \la \Y_{-E}| \J_1 \ra$} and
\hbox{$({d^2/d \ti^2} + W)$}
\hbox{$\la \Y^{j}_{W}| \R(\ti,\cdot) \ra$}
\hbox{$= \la \Y^{j}_{W}| \J_0 \ra + \ti \la \Y^{j}_{W}| \J_1 \ra$}
for \hbox{$W>0$}; these ODEs for the components of $\R(\ti,\cdot)$
are solved by elementary means, and one obtains:
\parn
\vbox{
    \beq \R(\ti,x) = \left[ \la \Y_{- E} | \R_0 \ra \cosh(\sqrt{E} \ti) +
    \la \Y_{- E} | \R_1 \ra {\sinh(\sqrt{E} \ti) \over \sqrt{E}} \right. \label{rsol} \feq
    $$ + \left. \la \Y_{-E}| \J_0 \ra {\cosh(\sqrt{E} \ti) -1 \over E} +
    \la \Y_{-E}| \J_1 \ra {\sinh(\sqrt{E} \ti) - \sqrt{E} \ti \over E^{3/2}} \, \right] \Y_{-E}(x) $$
    $$ + \sum_{j=1,2} \int_{0}^{+\infty} d W
    \left[ \la \Y^j_{W} | \R_0 \ra \cos(\sqrt{W} \ti) +
    \la \Y^j_{W} | \R_1 \ra {\sin(\sqrt{W} \ti) \over \sqrt{W}} \right. $$
    $$ + \left. \la \Y^j_{W}| \J_0 \ra {1 - \cos(\sqrt{W} \ti) \over W} +
    \la \Y^j_{W}| \J_1 \ra {\sqrt{W} \ti - \sin(\sqrt{W} \ti) \over W^{3/2}} \, \right] \Y^j_{W}(x)~. $$
}
This equation determines the function $\R$, which in turn appears
in the expressions \rref{q} \rref{psi} for $\Q,\Phi$.
\parn
As a converse of the above statements, let us consider functions
$\Q_i, \R_i \in C^\infty(\reali,\reali)$ such that
$\R_i, \J_i \in \Ee(\reali,\reali)$ for $i=0,1$,
where the $\J_i$'s are defined by
Eqs. \rref{ui} \rref{ji}.
Defining $\R$ and, subsequently, $\Q$,
$\Phi$ via Eqs. \rref{rsol} \rref{q} \rref{ui} \rref{psi}, one can
show the following:
\begin{itemize}
    \item[(a)] For each $\ti \in \reali$, the map
    $\R(\ti,\cdot): x \mapsto \R(\ti,x)$ is in $\Ee(\reali,\reali)$;
    the map $\ti \mapsto \R(\ti,\cdot)$ is $C^\infty$ from $\reali$ to
    $\Ee(\reali,\reali)$.
    \item[(b)] $\Q,\R,\Phi \in C^\infty(\reali^2,
    \reali)$.
    \item[(c)] $\Q_i$ and $\R_i$ are initial data, i.e., they are related
    to $\Q$ and $\R$ as in Eq. \rref{qri}.
    \item[(d)] $\Q,\R,\Phi$ fulfill the linearized Einstein equations
    (\ref{lit}-\ref{lite}).
\end{itemize}
\subsection{Linear instability of the EBMT wormhole.}
\label{ourinst}
This is proved showing that the linearized
Einstein equations have solutions diverging in the large $\ti$
limit. The simplest solution of this kind is obtained choosing the
initial data
\beq
\R_0(x) := \Y_{-E}(x),~~\R_1(x) := 0,~~
\Q_{0}(x) = {- 2 \Y_{-E} (x) \over (1 + x^2)^{{3 \over 2}}},~~
\Q_1(x) := 0 \, .
\feq
Then $\la \Y_{-E} | \R_0 \ra = 1$, $\la \Y^j_{W} | \R_0 \ra = 0$
and Eqs. \rref{ui} \rref{ji} give $\P_i=0, \J_i=0$ for $i=0,1$.
From here and from Eqs. \rref{rsol} \rref{q} \rref{psi} we get
\parn
\vbox{
    $$
    \R(\ti,x) = \Y_{-E}(x) \cosh(\sqrt{E} \ti), \qquad \Q(\ti,x)=-\frac{2
        \R(\ti,x)}{(1+x^2)^\frac{3}{2}}\,, $$
    \beq \Phi(\ti,x)=\frac{\R_{x}(\ti,x)}{\sqrt{1+x^2}} +\C\,.
    \label{sol}
    \feq
}
\noindent
Clearly, this solution diverges exponentially for $\ti \vain \pm
\infty$; the same feature appears in many associated geometrical
objects. Let us consider, for example, the scalar curvature $R$ of
the spacetime metric; substituting Eqs. \rref{sol} into Eq.
\rref{scalesp} (and using the relation $\HH \Y_{-E} = - E \Y_{-E}$,
i.e., $\Y_{-E,x x} =$ \hbox{$(E - 3/(1 +x^2)^2)$} $\times \Y_{-E})$ we get
$$
R = - {2 a^2 \over (a^2 + \ell^2)^2}
+ {4 \ep \over a^2}\K\left({\ell \over a}\right)
\cosh \left(\sqrt{E} \,{t \over a}\right) + O(\ep^2),
$$
\beq \K(x) := \left({1 \over (1 + x^2)^{7/2}}
- {E \over (1 + x^2)^{3/2}}\right) \Y_{-E}(x)
+ {x \over (1 + x^2)^{5/2}}\Y_{-E,x}(x)~.
\label{scalespsolle}
\feq
We remark that the above function $\K$ is not identically zero; in
particular ({\footnote{$1-E \neq 0$ due to the estimate for $E$ in
        \rref{defe}; let us show that $\Y_{-E}(0) \neq 0$. To this purpose
        let us recall that $\Y_{-E}$ is an even function, whence
        $\Y_{-E,x}(0)=0$; if it were also $\Y_{-E}(0)=0$, making obvious
        considerations on the initial value problem for the differential
        equation $\Y_{-E,x x} = (E - 3/(1 +x^2)^2) \Y_{-E}$ we could infer
        $\Y_{-E}(x) =0$ for all $x \in \reali$.}}),
\beq \K(0) = (1 - E) \Y_{-E}(0) \neq 0 ~. \feq
Let us also stress
that the divergence for $t \vain \pm \infty$ of the coefficient of
$\ep$ in Eq.\rref{scalespsolle} is not an artifact that one could
eliminate by an everywhere smooth coordinate change
$(t,\ell) \mapsto (\tp,\ellp)$, $\ep$-close to the identity.
In fact, let us consider any coordinate change of the form
\beq t = \tp + \ep a \T\left({\tp \over a},{\ellp \over a}\right) \, ,
\qquad \ell = \ellp + \ep a \L\left({\tp \over a},{\ellp \over a}\right)
\feq
where $\T,\L : \reali^2 \to \reali$ are smooth (dimensionless)
functions; then, Eq. \rref{scalespsolle} gives
\beq
R =  - {2 a^2 \over (a^2 + \ellp^2)^2} + {4 \ep \over a^2}
\Bigg[ \K\left({\ellp \over a}\right)
\cosh \left(\sqrt{E} \,{\tp \over a}\right)
+ { 2 (\ellp/a) \L(\tp/a, \ellp/a) \over (1 + \ellp^2/a^2)^3 }
\Bigg] + O(\ep^2). \label{scalespsolleps}
\feq
In particular, at spacetime points with $\ellp=0$ we have \beq R =
- {2 \over a^2} + {4 \K(0) \ep \over a^2} \cosh \left(\sqrt{E}
\,{\tp \over a}\right) + O(\ep^2) \, , \label{scalespsolleepsze}
\feq and the coefficient of $\ep$ in the above equation diverges
(again exponentially) for $\tau \to \pm \infty$, due to the
previous remark $\K(0) \neq 0$.
\vskip 0.4cm \noindent
{\large{\textbf{Acknowledgments}}}
This work was supported by: INdAM,
Gruppo Nazionale per la Fisica Matematica; INFN; MIUR, PRIN 2010
Research Project ``Geometric and analytic theory of Hamiltonian
systems in finite and infinite dimensions.''; Universit\`{a} degli
Studi di Milano.
\parn
We acknowledge K.A. Bronnikov, J.A.
Gonz\'{a}lez, F.S. Guzm\'{a}n, R.A. Konoplya and O. Sarbach for
encouragement, very useful exchange of views and bibliographical
references.


\begin{thebibliography}{99}
\bibitem{Mor} M.S. Morris and K.S. Thorne, \textsl{Wormholes in spacetime and their use for interstellar travel:
    A tool for teaching general relativity}, Am. J. Phys \textbf{56}
(1988), 395-412.
\bibitem{Div} O. James, E. von Tunzelmann, P. Franklin, and K.S. Thorne
\textsl{Visualizing Interstellar's Wormhole}, Am. J. Phys.
\textbf{83} (2015), 486-499.
\bibitem{Ell} H. Ellis, \textsl{Ether flow through a drainhole: A particle model in general relativity},
J. Math. Phys. \textbf{14} (1973), 104-118.
\bibitem{Bron73} K.A. Bronnikov, \textsl{Scalar-tensor theory and scalar charge},
Acta Phys. Polon. \textbf{B4} (1973), 251-266.
\bibitem{Yaz} S. Yazadjiev, \textsl{Uniqueness theorem for static wormholes in Einstein
    phantom scalar field theory}, Phys. Rev. D \textbf{96} (2017),
044045 (6 pp.).
\bibitem{Hawk} S.W. Hawking, G.F.R. Ellis, \textsl{The large scale structure
    of space-time}, Cambridge University Press, Cambridge (1973).
\bibitem{Vis} M. Visser, \textsl{Lorentzian wormholes. From Einstein to Hawking}, Springer-Verlag, New York (1996).
\bibitem{Proc} J.A. Gonz\'{a}lez, F.S. Guzm\'{a}n and O. Sarbach,
\textsl{On the instability of static, spherically symmetric
    wormholes supported by a ghost scalar field}, in: ``CP1083,
Gravitation and Cosmology, Proceedings of the Third International
Meeting'' edited by F. S. Guzman Murillo, A. Herrera-Aguilar, U.
Nucamendi and I. Quiros, American Institute of Physics (2008), pp.
208-216.
\bibitem{Gon} J.A. Gonz\'{a}lez, F.S. Guzm\'{a}n and O. Sarbach,
\textsl{Instability of wormholes supported by a ghost scalar
    field. I. Linear stability analysis}, Classical and quantum
gravity \textbf{26} (2009), 015010 (14 pp.).
\bibitem{Gon2} J. A. Gonz\'{a}lez, F. S. Guzm\'{a}n and O. Sarbach,
\textsl{Instability of wormholes supported by a ghost scalar
    field. II. Nonlinear evolution}, Classical and quantum gravity
\textbf{26} (2009), 015011 (20 pp.).
\bibitem{Bron2011} K. A. Bronnikov, J. C. Fabris and A. Zhidenko,
\textsl{On the stability of scalar-vacuum space-times}, The
European Physical Journal C-Particles and Fields \textbf{71}
(2011), 1791 (12 pp.).
\bibitem{Bron2018} K. A. Bronnikov, \textsl{Scalar fields as sources for wormholes and regular black holes},
Particles \textbf{2018}, 1, 56-81; doi:10.3390/particles1010005~.
\bibitem{Nov} A.A. Shatskii, I. D. Novikov, N. S. Kardashev,
\textsl{A dynamic model of the wormhole and the Multiverse model},
Physics Uspekhi \textbf{51} (5) (2008), 457 - 464.
\bibitem{Bronfl} K. A. Bronnikov, L. N. Lipatova, I. D. Novikov, A. A. Shatskiy,
\textsl{Example of a stable wormhole in general relativity},
Grav. Cosmol. \textbf{19} (4) (2013), 269-274.
\bibitem{Kon} R.A. Konoplya,
\textsl{How to tell the shape of a wormhole by its quasinormal
    modes}, Physics Letters B \textbf{784} (2018), 43-49.
\bibitem{Lan} L.D. Landau, E.M. Lifhsitz, \textsl{Course of Theoretical Physics, Vol. II:
    The classical theory of fields}, Fourth English Edition, Pergamon
Press, Oxford (1975).
\bibitem{Rud} W. Rudin, \textsl{Functional analysis}, 2nd Edition, McGraw-Hill, New York, 1991.
\bibitem{Ber} F.A. Berezin, M.A. Shubin, \textsl{The Schr\"odinger equation},
Mathematics and its Applications (Soviet Series) \textbf{66},
Kluwer Academic Publishers, Dordrecht (1991).
\end{thebibliography}
\end{document}